# Magnetic-field-induced topological phase transition in Fe-doped (Bi,Sb)$_2$Se$_3$ heterostructures

Y. Satake[1,2], J. Shiogai[1,*], G. P. Mazur[2], S. Kimura[1], S. Awaji[1], K. Fujiwara[1], T. Nojima[1], K. Nomura[1,3], S. Souma[3,4] T. Sato[3,4,5], T. Dietl[2,4], and A. Tsukazaki[1,3]

[1]*Institute for Materials Research, Tohoku University, Sendai 980-8577, Japan*

[2]*International Research Centre MagTop, Institute of Physics, Polish Academy of Sciences, Aleja Lotnikow 32/46, PL-02668 Warsaw, Poland*

[3]*Center for Spintronics Research Network (CSRN), Tohoku University, Sendai 980-8577, Japan*

[4]*WPI-Advanced Institute for Materials Research, Tohoku University, Sendai 980-8577, Japan*

[5]*Department of Physics, Tohoku University, Sendai 980-8578, Japan*

*Author to whom correspondence should be addressed.

Electronic mail: junichi.shiogai@imr.tohoku.ac.jp

**Abstract**

Three-dimensional topological insulators (3D-TIs) possess a specific topological order of electronic bands, resulting in gapless surface states via bulk-edge correspondence. Exotic phenomena have been realized in ferromagnetic TIs, such as the quantum anomalous Hall (QAH) effect with a chiral edge conduction and a quantized value of the Hall resistance $R_{yx}$. Here, we report on the emergence of distinct topological phases in paramagnetic Fe-doped $(Bi,Sb)_2Se_3$ heterostructures with varying structure architecture, doping, and magnetic and electric fields. Starting from a 3D-TI, a two-dimensional insulator appears at layer thicknesses below a critical value, which turns into an Anderson insulator for Fe concentrations sufficiently large to produce localization by magnetic disorder. With applying a magnetic field, a topological transition from the Anderson insulator to the QAH state occurs, which is driven by the formation of an exchange gap owing to a giant Zeeman splitting and reduced magnetic disorder. Topological phase diagram of $(Bi,Sb)_2Se_3$ allows exploration of intricate interplay of topological protection, magnetic disorder, and exchange splitting.

**I. Introduction**

Three-dimensional topological insulators (3D-TI) are a class of matter that composed of gapless surface states and insulating bulk [1-3]. In such material systems, various types of topological phase transitions (TPTs) are expected to appear as a result of structural modifications and application of external fields [4-13]. For instance, an exchange gap formation in the surface states engenders a TPT, which leads to the quantum anomalous Hall (QAH) phase, i.e., to the Chern insulator. In this phase, similarly to the conventional quantum Hall effect (QHE), charge transport proceeds via a dissipationless chiral edge channel while the origin is completely different [14-16]; the QAHE is driven by the formation of an exchange gap and the QHE by Landau level splitting. To materialize the QAH phase in TIs, the surface electronic states should be modified by the exchange gap owing to some mechanisms (1) coupling between surface state and ferromagnetically aligned spins of transition metal impurities [15-17] or aligned spins by external magnetic field in the antiferromagnetic topological insulator [18,19], (2) proximity coupling effect with adjacent magnetic moment [20-22], or (3) band crossing driven by a Zeeman splitting [17,23]. To date, experimental observations of the QAH effect have been accomplished for Cr-doped or V-doped $(Bi,Sb)_2Te_3$ (BST) 3D-TI films, where a perpendicular spontaneous magnetization develops below the Curie temperature (type (1)) [15,16,24-28], and for a proximitized (Bi,Sb)$_2$Te$_3$/ferromagnetic insulator heterostructure (type (2)) [29]. In contrast, type (3) of QAH has not been observed in paramagnetic (Bi,Sb)$_2$Se$_3$-based TIs, whereas the quantization of the Hall resistance is accomplished by the application of a magnetic field to a 4 quintuple layer of Cr-doped (Bi,Sb)$_2$Te$_3$[30], Ti-doped (Bi,In)$_3$Te$_2$ [31], and MnBi$_2$Te$_4$ [18,19].

As another TPT, the surface state can be gapped when the 3D-TI film is sufficiently thin to activate hybridization of states from the opposite surfaces [1,2]. Thin films of Bi$_2$Se$_3$-type 3D-TIs are known to be 2D trivial insulators [4,32-34] when the hybridization is sufficiently strong, as presented in Fig. 1(a). Considering this trivial electronic band as an initial state, the giant Zeeman splitting generated by an external magnetic field produces a specific band crossing (Fig. 1(b)) with formation of an exchange gap (Fig. 1(c)) [17,23] at this crossing point, equivalent to that driven by type (1) and (2) in coupling with ferromagnetic systems, inducing a conceptually different type (3) of QAH. To illustrate the topological phase transition from the hybridized insulator to the QAH phase driven by the Zeeman splitting, we have calculated the band structure using a 3D-TI slab geometry. Figures 1(d)-1(f) demonstrate the band modification of 3D-TI by the inter-surface hybridization and the Zeeman effect. When $\Delta_{\text{hy}}$ is larger than $\Delta_{\text{Zeeman}}$, the hybridized insulator is stabilized without surface gapless states (Fig. 1(d)).

The $Δ_{Zeeman}$ exceeding $Δ_{hy}$ forms a non-trivial exchange gap with one-dimensional chiral-edge channel as shown in Figs. 1(e) and 1(f), with gap size proportional to the magnetic field. In Fig. 1(f), red (blue) thick curve shows the chiral edge mode at one (the other) edge.

With a minor contribution of the ordinary Hall effect under magnetic field owing to Fermi energy ($E_F$) locating in the gap, the magnetic-field-induced anomalous Hall effect plays a critical role for the quantization of the Hall conductivity. This study demonstrates the presence of TPTs between 3D-TI, a two-dimensional (2D) insulator, and QAH phases in paramagnetic Fe-doped $Bi_2Se_3$-based heterostructures by controlling Fe doping and layer thickness and by application of external magnetic and electric fields.

The tetradymite compound $Bi_2Se_3$ is a representative 3D-TI, in which a large bulk band gap (approximately 300 meV) hosts gapless surface states [35]. Using angle-resolved photoemission spectroscopy (ARPES) of surface states, 50 meV gap formation was detected in paramagnetic Fe-doped bulk $Bi_2Se_3$; it was assigned to breakage of the time-reversal symmetry by exchange interactions [36]. Effects of magnetic impurities upon topological surface states were also studied by *in-situ* deposition of Fe atoms [37-39]. It was found that Fe acts as donor [37,38] if deposited at room temperature but as an acceptor if deposited at 8 K [38]. A question on whether the presence of a Fe surface layer opens a gap or not in the topological states was also examined experimentally and theoretically [37-39].

According to one theoretical proposal, Fe-doped $Bi_2Se_3$-based heterostructures are a preferred platform for observing the emergence of QAH phase [17]. Nevertheless, no QAH effect has yet been observed in magnetically doped $Bi_2Se_3$-based films, probably because of the absence or weakness of the ferromagnetic ordering, and associated difficulties in tuning of the $E_F$ into a correspondingly small exchange gap [40]. In contrast, a TPT from the TI to the 2D-insulator by film thickness reduction was demonstrated for $Bi_2Se_3$ films using ARPES and electrical measurements [4,31-33]. Depending on the experimental method used, the critical thickness for the transition has been reported as 5 [4] or 10 nm [41] in $Bi_2Se_3$. For other 3D-TIs such as $Bi_2Te_3$, $(Bi,Sb)_2Te_3$ (BST) and magnetically doped BST films, variations in the strength of the spin-orbit interaction are thought to account for critical thickness changes from 1 to 13 nm [12,42].

## II. EXPERIMENTAL DETIALS

Employing molecular-beam epitaxy (MBE) and a 3-nm-thick $n$-type $Bi_2Se_3$ buffer layer on a semi-insulating Fe-doped InP (111) substrate, we have grown $Fe_x(Bi_{1-y}Sb_y)_{2-x}Se_3$ films of thickness $d$ ranging from 8 to 30 nm (see Supplementary section 1 [43]),. The control parameters are the Sb composition and the gate bias voltage for changing the $E_F$ position [44,45], whereas the Fe concentration and the external magnetic field serve for tuning magnetic disorder and band splitting state. Because Fe doping up to at least $x = 0.1$ (nominal composition controlled by the beam flux ratio) produces rather minor changes in $R_{xx}$ and $|R_{yx}|$ magnitudes (see Supplementary section 2 [43]), we infer that Fe ions are isoelectronic impurities occupying Bi or Sb sites, and assume the same $Fe^{3+}$ charge configuration with bulk Fe-doped $Bi_2Se_3$ [36] that typically corresponds to the high spin $S = 5/2$. However, Fe acts not only as a magnetic impurity, but also as an isoelectronic dopant, in a manner similar to Sb, which affects the carrier concentration owing to band shifts with respect to defect formation levels. Moreover, thickness $d$ plays a crucially important role in the control of both (i) hybridization between surface and interfacial topological states and (ii) the width of the depletion layer in the *p-n* junction that forms at the interface to the *n*-type $Bi_2Se_3$ buffer layer.

## III. RESULT AND DISCUSSION

**A. Angle-resolved photoemission spectroscopy (ARPES) and magnetic properties**

We have characterized the electronic structure of a 30-nm-thick $Fe_{0.05}(Bi_{0.34}Sb_{0.66})_{1.95}Se_3/Bi_2Se_3$ heterostructure at 40 K using ARPES. As presented in Fig.

1(g), the topological surface states and the bulk valence band are resolved clearly at the Γ point, indicating that this thick Fe-doped $(Bi,Sb)_2Se_3$ layer preserves key features of 3D-TIs and does not engender a substantial Fermi level shift. Linear dispersion of the surface band is more readily apparent when moving the Fermi level up in energy by an aging process that increases the surface electron concentration (see Fig. S3 [43]). Contrary to an earlier report on ARPES at 10 K of bulk Fe-doped $Bi_2Se_3$ samples with Fe concentration $x = 0.12$ and $0.16$, which was considered as a consequence of ferromagnetic ordering [36], no sizable gap in the surface states is resolved in the case of our MBE-grown layer with $x = 0.05$; this indicates the absence of a long range magnetic order at Fe concentration of interest here.

The ARPES data are consistent with the magnetotrasport studies: neither hysteretic nor saturation behavior has been observed in our Hall effect measurements, indicating that Fe ions remain in a paramagnetic phase down to 2 K for $x = 0.05$. This is in agreement with previous direct measurements of magnetization $M(T,H)$ down to 2 K and up to $\mu_0 H = 5$ T on bulk samples that showed paramagnetic behavior below $x = 0.16$ [36]. Some hysteresis superimposed on the paramagnetic signal $M(H)$ were detected for $x \geq 0.16$ at 2 K [36]. The absence of ferromagnetism at low Fe concentrations is consistent with our direct magnetization measurements employing superconducting quantum interference device (SQUID), which do not show any hysteresis. However, a quantitative evaluation of the epilayer paramagnetic signal from our SQUID and electron paramagnetic resonance studies has been hampered by a small thickness of the epilayers and a large magnetic contribution of the Fe-doped InP substrate. The absence of ferromagnetism in the Fe-doped $Bi_2Se_3$ is also consistent with *ab initio* studies [46,47].

## B. Magnetotransport properties

The values of the 2D resistivity tensor components, *i.e.*, the sheet resistance $R_{xx}$ and the Hall resistance $R_{yx}$, have been measured in the Hall-bar geometry and using a standard lock-in technique [43]. Figure 1(h) presents a contour plot of $R_{xx}$ values for $Fe_{0.05}(Bi_{0.34}Sb_{0.66})_{1.95}Se_3/Bi_2Se_3$ films in $B = 0$ as a function of layer thickness $d$ and temperature $T$ (see Supplementary Fig. S4(a) for detailed dataset [43]). One phenomenological criterion describing the critical point of the localization transition in the 2D case is the sheet resistance value of $h/e^2 \cong 25.8$ k$\Omega$ [48,49]. The boundary is pronounced as a white color region in Fig. 1(h). As shown there, $R_{xx}(T)$ determined for the 30-nm-thick film shows lower values than that of $h/e^2$, in addition to weak temperature dependence, implying that this sample is the 3D-TI phase (Fig. 1(g)). With decreasing $d$ at fixed Fe and Sb concentrations ($x = 0.05$ and $y = 0.66$), $R_{xx}(T)$ reaches the critical value of $h/e^2$ at thickness of $d = 14$ nm, below which dependence $R_{xx}(T)$ exhibits insulating behavior and $R_{xx}$ magnitudes exceed 1 MΩ at the lowest temperature of 2 K (deep-red region in Fig. 1(h)). Such highly insulating features at $d < 14$ nm can be ascribed either to the trivial Anderson insulator [50,51] or to the quantum spin Hall insulator [51,52], if the formation of edge states would accompany hybridization of surface states. However, edge conductance would hardly dominate the charge transport in the present Hall-bar geometry with the channels longer than the expected protection length of helical topological channels. The phase coherence length estimated by weak antilocalization analysis of the $Bi_2Se_3$ thin films is about 200 nm [45], which is well below the channel length. In principle, ARPES is an effective way to tell the topological phase. However, the high resistance and the corresponding charging effect made it difficult to assess the electronic structure of our thinnest samples. Nevertheless, the critical thickness we found for this 2D-insulator is rather thick compared to the values (5–10 nm) reported for non-magnetic $Bi_2Se_3$ films [4,41]. This point can be explained by the empirical fact (see Supplementary Section 5 [43]) that the substitution of heavier Bi or Sb by lighter Fe reduces the spin-orbit interaction strength, which might increase the critical thickness. Furthermore, the presence of the $n$-type $Bi_2Se_3$ buffer layer depletes the $p$-type $Fe_{0.05}(Bi_{0.34}Sb_{0.66})_{1.95}Se_3$ layer [45], making the effective thickness less than the nominal value.

More importantly, in our paramagnetic samples, spin disorder scattering by Fe impurities can be effective for expanding the Anderson insulator range with regard to the thickness and temperature of $Fe_x(Bi_{1-y}Sb_y)_{2-x}Se_3$ films (Fig. 1(h)), making it wider compared to $Bi_2Se_3$. Temperature dependence of resistivity for samples with $d = 14$ nm and various $x$ are presented in Fig. S2 [43]. The data show clearly a transition from the weakly localized regime to the strongly localized regime, characterized by a sharp increase of resistivity on lowering temperature, which appears at $x > 0.04$. As expected in the absence of Landau quantization and in the 2D case [53], this transition has a crossover character, so that it does not obey scaling equations. As presented in Fig. 1(i), the magnitude of magnetoresistance defined as MR = $[R_{xx}(B) - R_{xx}(0)]/R_{xx}(0)$ exceeds −80% in 9 T and at 2 K in the vicinity of the localization transition. The ordering of Fe spins in the magnetic field reduces magnetic disorder and enhances the spin splitting of electronic states. Consequently, a QAH phase would appear in the thickness region toward type (3) by the application of the magnetic field as schematically shown in Fig. 1(c) and 1(f). In accordance with the expectation that reduction of magnetic disorder occurs for any field direction, we find negative MR to be also present for the in-plane magnetic field (Supplementary Fig. S6 [43]).

Figure 2 presents a dependence of Hall resistance $R_{yx}$ on the Sb concentration $y$

at $T = 2$ K and $B = 9$ T for $Fe_{0.05}(Bi_{1-y}Sb_y)_{1.95}Se_3/Bi_2Se_3$ with $d \cong 14$ nm (red circles). This dependence is strikingly different compared to the case of non-magnetic $(Bi_{1-y}Sb_y)_2Se_3/Bi_2Se_3$ films with $d = 14$–$20$ nm (black squares) [44]. The representative raw data of $R_{yx}(B)$ for Fe-doped samples are shown in the inset. For non-magnetic $(Bi_{1-y}Sb_y)_2Se_3/Bi_2Se_3$, the sign reversal of $R_{yx}$ reflects the conversion of the carrier type and therefore the tuning of $E_F$ across the charge neutrality point (CNP), which appears at $y \cong 0.7$. In contrast, the $R_{yx}$ in the Fe-doped films persists positive value in $y > 0.5$ indicates that the positive contribution of magnetic-field-induced anomalous Hall resistance is much larger than the negative contribution of ordinary Hall component; empirically $R_{yx} = R_0 B + R_A M_z$ where $R_0$ is the ordinary Hall coefficient, $B$ magnetic field, $R_A$ the anomalous Hall coefficient, and $M_z$ out-of-plane component of magnetization. By application of a perpendicular $B$, $B$–induced $M_z$ is dominantly contribute to the $R_{yx}$ with negligibly small ordinary component. Moreover, $R_{yx}$ is enhanced greatly with a peak at $y \cong 0.67$, close to the CNP in non-magnetic films. The behavior of $R_{yx}$ as a function of $y$, together with a large magnitude of $R_{yx}$ at the CNP, demonstrate that $R_{yx}$ is dominated by the intrinsic anomalous Hall effect, as its amplitude is expected to be dependent on the $E_F$ position while its sign is determined by magnetization direction and exchange coupling [17,19,54]. In fact, the tangent of Hall angle $R_{yx}/R_{xx}$ approaches unity at values greater than $d$ = 14 nm (see Supplementary Fig. S4(f) [43]) indicating that the condition for the Landau quantization of the density of states, $R_0B/R_{xx}$ >1, is not met. It should be noted that such a large $R_{yx}$ at $d$ = 14 nm is observed also in the insulator region (Fig. 1(h)), where negative MR exists (Fig. 1(i)).

Together with revealing enhanced values of $R_{yx}(B)$, transition behavior is found in $R_{xx}(T)$ in elevated fields up to 24 T, as depicted in Fig. 3(a). The field-induced insulator-metal transition becomes apparent below 50 K. For $B$ > 12 T, the metallicity increases concomitantly with increasing $B$ down to temperatures as low as 1.6 K. The criterion in $B$ = 9 T for the sign change of d$R_{xx}$/d$T$ being close to 0.5 $h/e^2$. This value for the MIT, compared to 1 $h/e^2$ for MIT driven by structural modifications (cf. Fig. 1(h)), indicates one other type of the TPT. We attributed this field-induced metallization to a TPT from the 2D-insulator to a QAH phase. The insulator at $B$ = 0 with hybridization gap $\Delta_{hy}$ (Fig. 1(a) and 1(d)) turns to be a state with inverted bands by Zeeman splitting $\Delta_{Zeeman}$ when $\Delta_{Zeeman}$ > $\Delta_{hy}$. With the assistance of spin-orbit coupling, a gap is formed at the crossings of surface bands hosting carriers with opposite spin orientations [17,30], as depicted in Fig. 1(c) and 1(f). Furthermore, at $\Delta_{Zeeman}$ > $\Delta_{hy}$, gapless disorder-protected chiral channels are formed, in a full analogy to the theoretical proposal for the QAH effect [17].

Having observed the signature of this new TPT, we study $R_{xx}(B)$ and $R_{yx}(B)$ at $T$

= 1.6 K employing a field-effect transistor (FET) device for precise tuning of $E_F$. The FET layout is depicted in the inset shown with Fig. 3(b) (see also Supplementary Materials [43]) – it consists of a 1-nm-thick $Bi_2Se_3$/18.5-nm-thick $Fe_{0.05}(Bi_{0.33}Sb_{0.67})_{1.95}Se_3$/3-nm-thick $Bi_2Se_3$ trilayer structure. As depicted in Fig. 3(b), by sweeping magnetic field up to 15 T under $V_G = -60$ V, the large $R_{xx}$ value of 3.92 $h/e^2$ in $B = 0$ (black line) decreases monotonically to 0.2 $h/e^2$ whereas $R_{yx}$ (red line) increases, reaching 0.975 $h/e^2$, i.e., almost the value expected for the ideal QAH phase, $h/e^2$. Consequently, the magnetic-field-induced transition from the 2D-insulator to the QAH phase can be observed clearly at the high temperature of $T = 1.6$ K, at which the quantization accuracy of the $R_{yx}$ value is comparable to that reported for Cr modulation doped $(Bi,Sb)_2Te_3$ heterostructures [15,55] and higher than that for Cr-doped or V-doped $(Bi,Sb)_2Te_3$ thin films [8,24-28]. Note that this quantization is hardly considered to be QHE due to insulating initial state at $B = 0$ with a negligible ordinary Hall effect. The electrostatic tuning of the $E_F$ position with respect to the gap around the CNP promotes other TPTs as presented in conductivity tensor in Supplementary Fig. S10 [43].

The nature of relevant phases is assessed by examining the renormalization group flow of the conductivity tensor components [$\sigma_{xx} = R_{xx}/(R_{xx}^2 + R_{yx}^2)$, $\sigma_{xy} = R_{yx}/(R_{xx}^2 + R_{yx}^2)$] as a function of temperature [8,25,56]. Figures 4(a) and 4(b) show $\sigma_{xx}(T)$ and $\sigma_{xy}(T)$ for various $B$ at $V_G = -60$ V. As $T$ decreases, all $\sigma_{xx}(T)$ values decrease, irrespective of $B$. However, $\sigma_{xy}(T)$ goes to $e^2/h$ and to zero, respectively, in high and low fields, with a crossover point of about 0.5 $e^2/h$. These data are shown in the [$\sigma_{xy}(B)$, $\sigma_{xx}(B)$] plane in Fig. 4(c). By lowering $T$ from 15 K (open symbols) to 1.6 K (close symbols) under various magnetic fields, the data converge to either ($\sigma_{xx}$, $\sigma_{xy}$) = (0, 0) for the insulator or (0, $e^2/h$) for QAH phase. The finding of a converging point in our Fe-doped (Bi,Sb)$_2$Se$_3$ samples clearly evidences the presence of a phase transition driven by the magnetic field, with the critical field of 7–8 T corresponding to the point at which $\Delta_{Zeeman}$ becomes comparable to $\Delta_{hy}$. Moreover, as shown in Supplementary Fig. S10 [43], the transition driven by an electric field in fixed magnetic fields of 9 and 12 T is visible. Given the experimental observation of the TPT in the accessible magnetic field, the large $g$-factor of the surface state should be taken into consideration to estimate $\Delta_{Zeeman}$, as discussed in Supplementary Section 9 [43]. The Zeeman splitting ($g$-factor of Bi$_2$Se$_3$ ~ 18-50 [57,58]) can be enhanced considerably over the band value $g^*\mu_B B$ through the $sp$-$d$ exchange interaction between itinerant carriers and localized magnetic moments, as observed in various dilute magnetic semiconductors, primarily doped with Mn but also with Fe$^{3+}$ ions [59]. By applying $g$-factor engineering, the TPT might be shifted to much weaker magnetic fields and higher temperatures. The renormalization group flow for type (3) in paramagnetic (Bi,Sb)$_2$Se$_3$ film (Fig. 1(c) and 1(f)) can be understood in terms of universal QAH phenomenon as previously discussed in type (1) in ferromagnetic (Bi,Sb)$_2$Te$_3$ films [8,25,56].

**IV. Conclusion**

We have performed comprehensive study of topological phase transitions based on Fe-doped (Bi,Sb)$_2$Se$_3$ thin films, which are driven by structural modification, doping, and external magnetic and electric fields. We found that topological phase transition occurs from 3D-TI to 2D-insulator by Fe doping owing to localization by magnetic disorder. The 2D-insulator is turned to QAH state by external magnetic field owing to a large Zeeman splitting and suppression of the magnetic disorder. Demonstration of the magnetic-field-induced TPT toward the QAH state in Bi$_2$Se$_3$-based 3D-TI sheds light on new perspectives of the QAH physics and expands it to a broader class of paramagnetic TI materials with properties that can be controlled by *g*-factor engineering.

**Acknowledgments**

This work was partly supported by CREST (No. JPMJCR18T1 and JPMJCR18T2), the Japan Science and Technology Agency, a Grant-in-Aid for Scientific Research on Innovative Areas (No. JP15H05853), and a Grant-in-Aid for Young Scientists (A) (No. 16H05981). Y.S. was supported by Kato Foundation for Promotion of Science (Grant No. KS-2914). The work in Poland was funded by the Foundation for Polish Science through the IRA Programme financed by EU within SG OP Programme. We thank NEOARK Corporation for the use of photolithography equipment for device fabrication.

**Figure legends**

FIG. 1 Schematic and calculated electronic band structure. (a-c) Evolution of surface spin subband structure in paramagnetic Fe-doped (Bi,Sb)$_2$Se$_3$ thin films upon thickness reduction and external magnetic field application. (a) Surface states of spin-up (magenta) and spin-down (cyan) subbands with gap $\Delta_{hy}$ driven by inter-surface hybridization, leading to 2D insulator phase. (b) Under an external perpendicular magnetic field, the spin subband degeneracy is lifted by the Zeeman effect, producing band crossing when the Zeeman energy $\Delta_{Zeeman}$ exceeds $\Delta_{hy}$. (c) At crossings of spin up (magenta) and spin down (cyan) bands, a gap is generated by the spin-orbit interaction, leaving a one-dimensional chiral edge channel in the bulk bandgap (red solid line). (d-f) Calculated band structure of three-dimensional topological insulator thin films. The band structure is calculated using the 3D-TI slab for cases in which (d) $\Delta_{hy}$ is larger than $\Delta_{Zeeman}$, (e) $\Delta_{hy}$ is comparable to $\Delta_{Zeeman}$, and (f) $\Delta_{Zeeman}$ is larger than $\Delta_{hy}$. (g) ARPES intensity of the 30-nm-thick Fe$_{0.05}$(Bi$_{0.34}$Sb$_{0.66}$)$_{1.95}$Se$_3$/3-nm-thick Bi$_2$Se$_3$ film around the $\overline{\Gamma}$ point measured at $T = 40$ K. SS and VB respectively represent surface state and bulk valence band. $E_F$ is indicated by a white dashed line. (h) Longitudinal resistance $R_{xx}$ of Fe$_{0.05}$(Bi$_{0.34}$Sb$_{0.66}$)$_{1.95}$Se$_3$/Bi$_2$Se$_3$ at $B = 0$ as a function of the magnetic layer thickness $d$ and temperature $T$. Colour bar scale is in the units of $h/e^2$. (i) Magnetoresistance of $Fe_{0.05}(Bi_{0.34}Sb_{0.66})_{1.95}Se_3/Bi_2Se_3$ at 2 K as a function of $d$ and $B$. Colour scale bar corresponds to +20 (blue) to –100 (red) %.

FIG. 2 Hall resistance of Fe-doped and non-magnetic heterostructures against Sb composition $y$ at 9 T and 2 K. Hall resistances $R_{yx}$ of 14-nm-thick $Fe_x(Bi_{1-y}Sb_y)_{2-x}Se_3$/3-nm-thick $Bi_2Se_3$ (red circles) dominated by the anomalous component and of 14–20-nm-thick non-magnetic $(Bi_{1-y}Sb_y)_2Se_3$/3-nm-thick $Bi_2Se_3$ films (black squares) are shown against Sb content $y$. CNP denotes the charge neutral point of non-magnetic samples. The inset shows the Hall effect measurement for $Fe_x(Bi_{0.33}Sb_{0.67})_{2-x}Se_3$/3-nm-thick $Bi_2Se_3$ films for Fe content $x = 0$ (solid black line), 0.04 (orange), and 0.05 (red).

FIG. 3 Magnetic-field-induced insulator-metal transition and quantum anomalous Hall effect. (a) Longitudinal sheet resistance $R_{xx}(T)$ of 14-nm-thick $Fe_{0.05}(Bi_{0.33}Sb_{0.67})_{1.95}Se_3/Bi_2Se_3$ heterostructure (inset) in various magnetic fields $B$. (b) Hall and sheet resistances, $R_{yx}(B)$ (solid red line) and $R_{xx}(B)$ (solid black line), measured for a FET device (shown in the inset) at $V_G = -60$ V and $T = 1.6$ K. The $R_{yx}$ data below 1 T were not obtained owing to the large value of $R_{xx}$.

FIG. 4 Magnetic-field-induced insulator-QAH phase transition. (a,b) Temperature dependences of the longitudinal and Hall conductivities ($\sigma_{xx}$ and $\sigma_{xy}$) at $V_G = -60$ V for various magnetic fields $B$. (c) Renormalization group flow in the [$\sigma_{xx}(T)$, $\sigma_{xy}(T)$] plane in various $B$ from 3 to 15 T. Each curve is extracted from data in (a) and (b) employing the same colour code. Empty and filled circles respectively present data obtained at $T = 15$ and 1.6 K. The [$\sigma_{xx}$, $\sigma_{xy}$] flow extracted from the data as a function of the magnetic field at $T = 1.6$ K is also shown (solid black line).

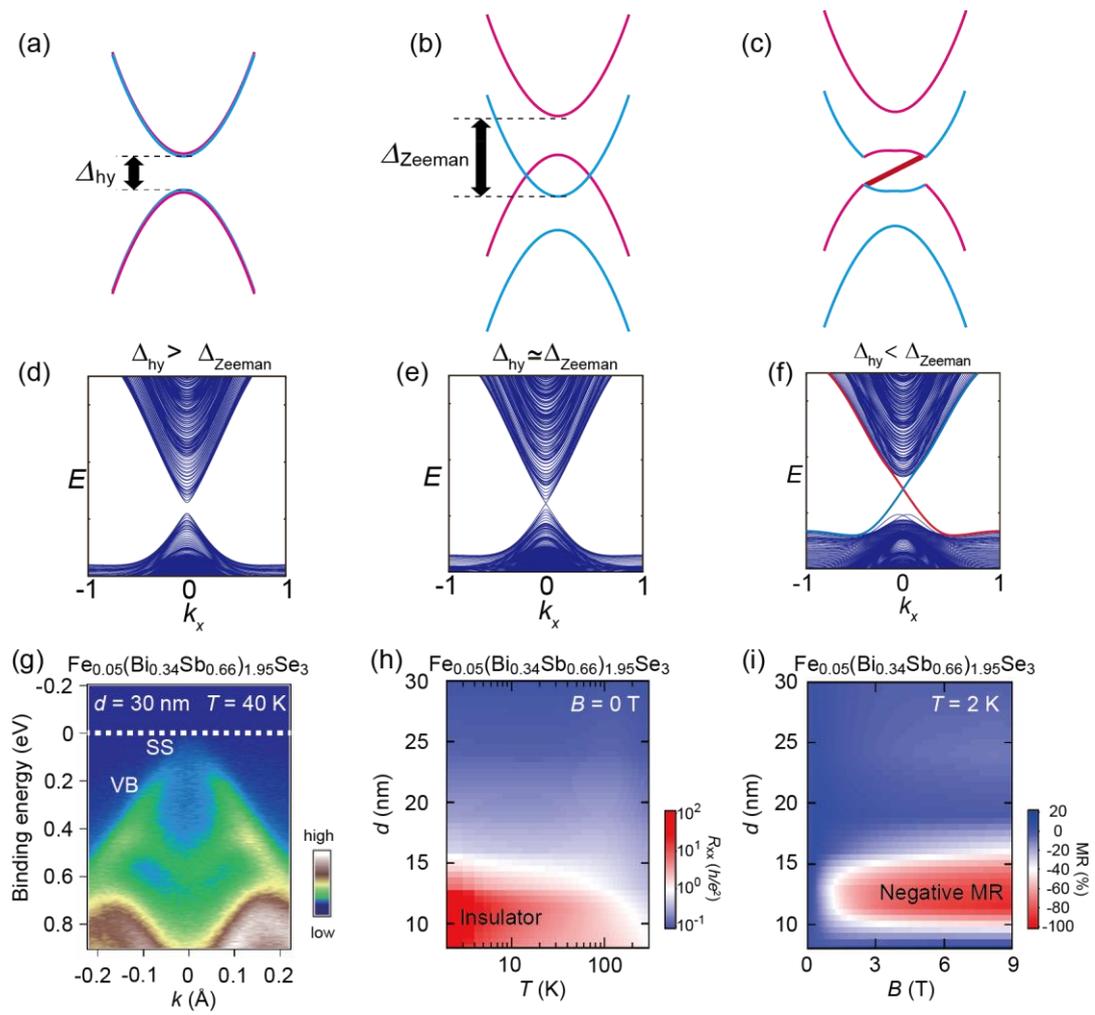

Figure 1 Y. Satake *et al.*

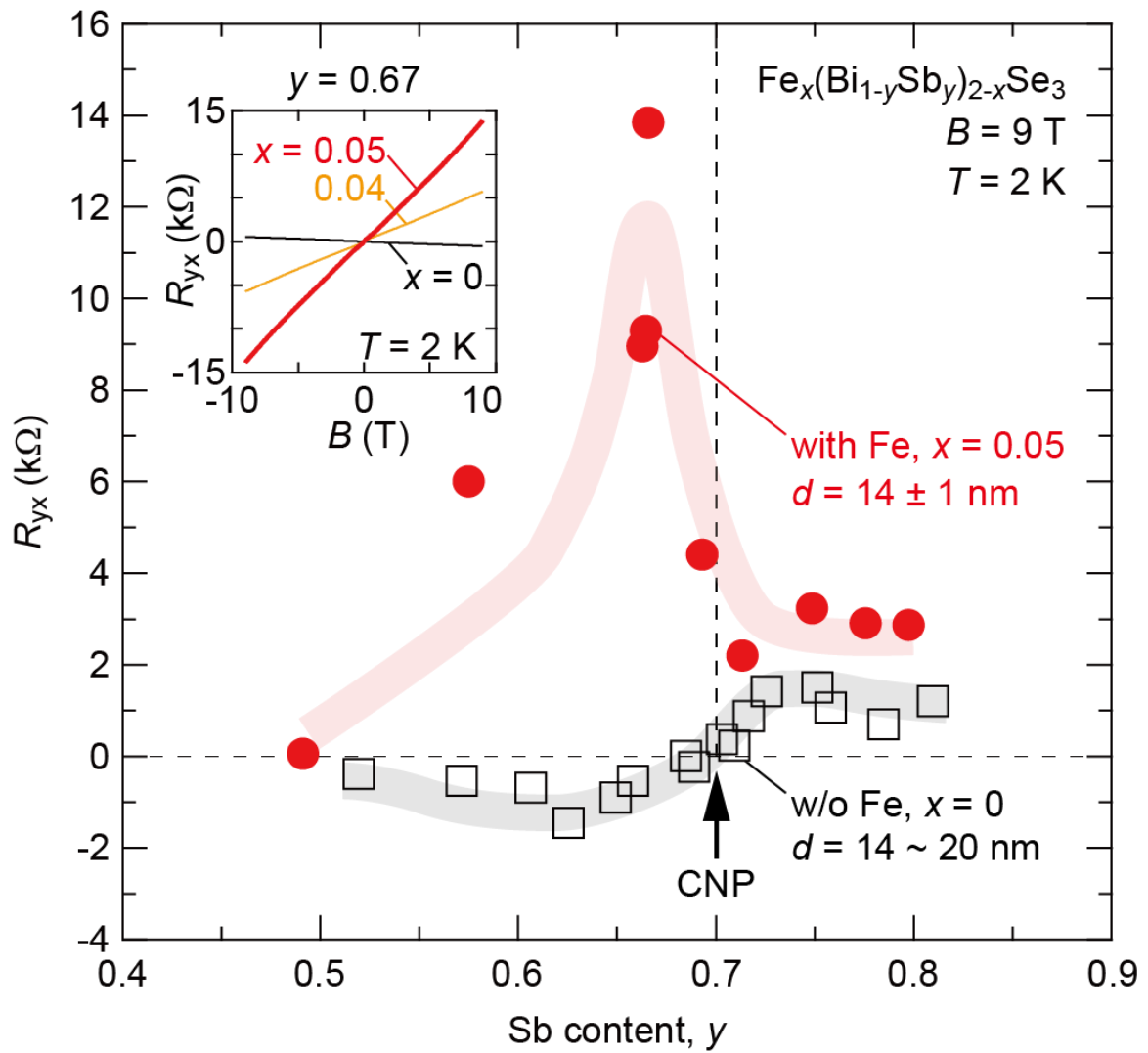

Figure 2 Y. Satake *et al*.

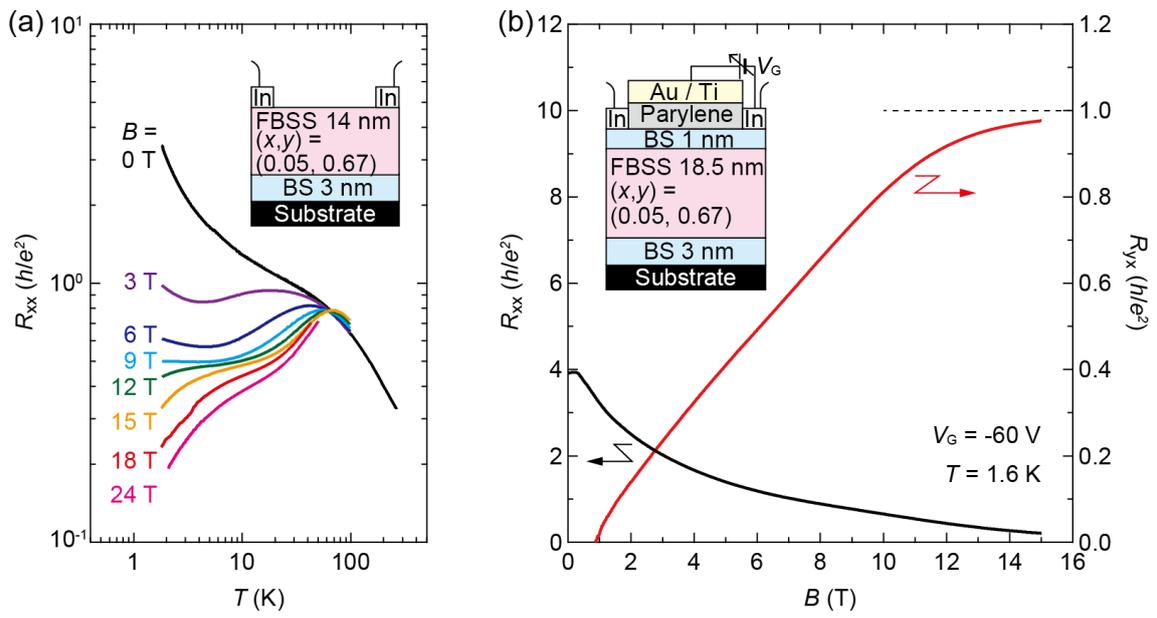

Figure 3 Y. Satake *et al.*

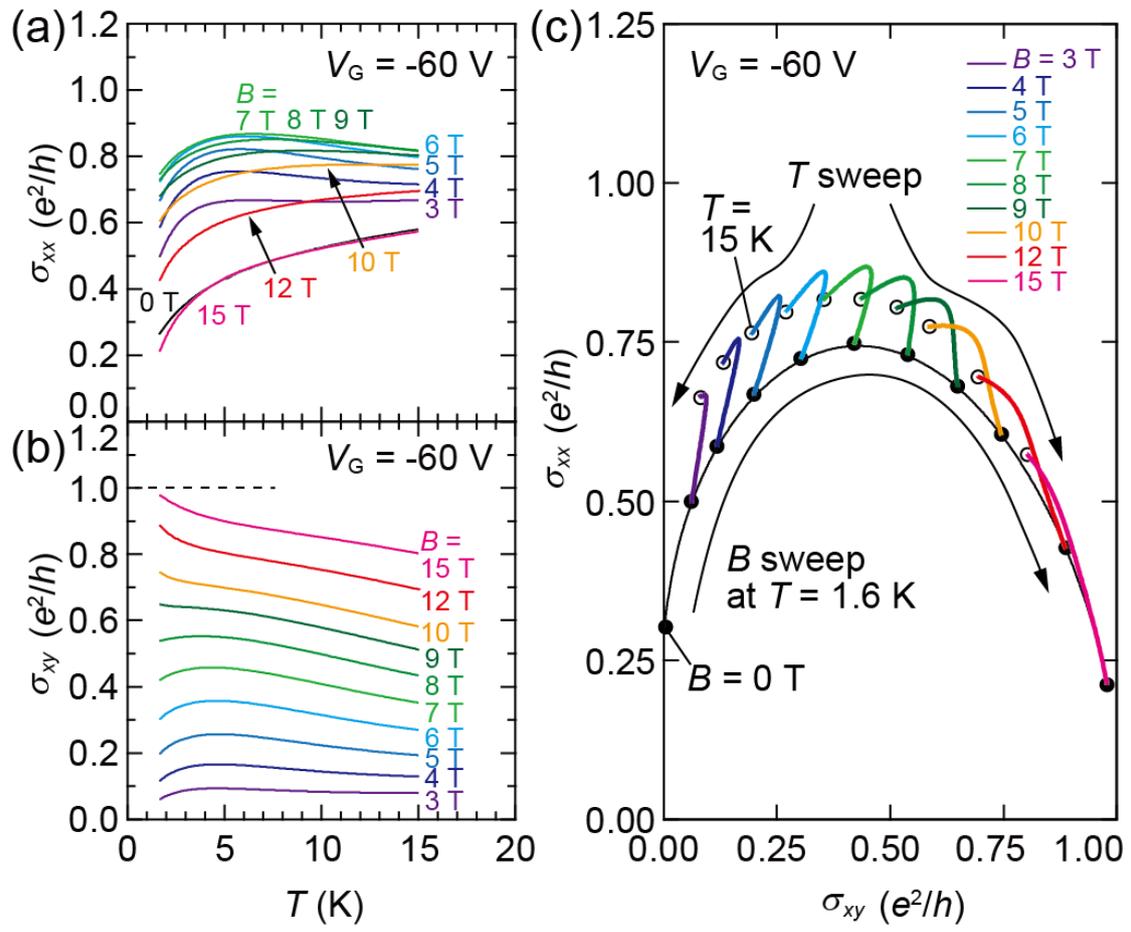

Figure 4 Y. Satake *et al.*

# Supplementary Materials for
# Magnetic-field-induced topological phase transition in Fe-doped (Bi,Sb)$_2$Se$_3$ heterostructures

Y. Satake[1,2], J. Shiogai[1*], G. P. Mazur[2], S. Kimura[1], S. Awaji[1], K. Fujiwara[1], T. Nojima[1], K. Nomura[1,3], S. Souma[3,4], T. Sato[3,4,5], T. Dietl[2,4] and A. Tsukazaki[1,3]

[1]*Institute for Materials Research, Tohoku University, Sendai 980-8577, Japan*

[2]*International Research Centre MagTop, Institute of Physics, Polish Academy of Sciences, Aleja Lotnikow 32/46, PL-02668 Warsaw, Poland*

[3]*Center for Spintronics Research Network (CSRN), Tohoku University, Sendai 980-8577, Japan*

[4]*WPI-Advanced Institute for Materials Research, Tohoku University, Sendai 980-8577, Japan*

[5]*Department of Physics, Tohoku University, Sendai 980-8578, Japan*

*Author to whom correspondence should be addressed.
Electronic mail: junichi.shiogai@imr.tohoku.ac.jp

Contents

# 1. Experimental details

## 1-1. Thin-film growth

Heterostructures consisting of $Fe_x(Bi_{1-y}Sb_y)_{2-x}Se_3$ layers with various thicknesses $d$ deposited onto 3-nm-thick $Bi_2Se_3$ buffer layer and trilayer structures consisting of 1-nm-thick $Bi_2Se_3$/18.5-nm-thick $Fe_x(Bi_{1-y}Sb_y)_{2-x}Se_3$/3-nm-thick $Bi_2Se_3$ were synthesized using molecular-beam epitaxy (MBE) on semi-insulating InP(111) substrates at 250°C. The flux of each element is characterized by the beam equivalent pressure. During the growth, Bi and Se fluxes were fixed at $3.5 \times 10^{-6}$ Pa and $1.7 \times 10^{-4}$ Pa, respectively, whereas Fe and Sb fluxes were varied depending on the desired contents. The Sb composition is calibrated using electron energy dispersive x-ray spectroscopy (EDX) and Rutherford back scattering (RBS) [1]. The Fe composition is estimated using the flux ratio. We regard the most favorable site of Fe as cation substitution [2,3]. Film thicknesses were ascertained from the Laue thickness fringes in x-ray diffraction patterns.

## 1-2. ARPES measurements

The films were transferred from the MBE chamber to the ARPES apparatus immediately after their growth. To transfer samples without breaking ultrahigh vacuum, we use a portable chamber that maintains ultrahigh vacuum. The base pressure of the portable chamber was kept below $7 \times 10^{-7}$ Pa. Measurements were taken at Tohoku University with a MBS A1 electron analyzer using a high-flux He discharge lamp and a toroidal grating monochromator. The He Iα ($h\nu = 21.218$ eV) line was used to excite photoelectrons.

## 1-3. FET device fabrication

The sample was patterned into Hall-bar geometry using standard photolithography and wet chemical etching. After etching, gate dielectric of 180-nm-thick parylene was deposited at room temperature, followed by the deposition of Au (100 nm)/Ti (10 nm) as a gate electrode. The channel width and center-to-center separation between longitudinal voltage probes were both approximately 100 μm. We adopted 1-nm-thick $Bi_2Se_3$/18.5-nm-thick $Fe_x(Bi_{1-y}Sb_y)_{2-x}Se_3$/3-nm-thick $Bi_2Se_3$ trilayer structure as a channel of the FET device. We expect that the top undoped $Bi_2Se_3$ capping layer suppresses disorder on the top of Dirac surface states possibly induced by Fe doping. Furthermore, the nearly symmetric structure consisting of 1-nm-thick $Bi_2Se_3$/18.5-nm-thick $Fe_x(Bi_{1-y}Sb_y)_{2-x}Se_3$/3-nm-thick $Bi_2Se_3$ trilayer cancels the energy differences of the Dirac points formed on the top and bottom surface states and depletes bulk carriers. Under the assumption that the number of hole carriers in 14-nm-thick $Fe_{0.05}(Bi_{0.33}Sb_{0.67})_{1.95}Se_3$ is almost equal to that of the electrons in 3-nm-thick $Bi_2Se_3$ bilayer, we increased the thickness of $Fe_{0.05}(Bi_{0.33}Sb_{0.67})_{1.95}Se_3$ to 18.5 nm to compensate for the effect of excessive electrons brought by deposition of the top 1-nm-thick $Bi_2Se_3$.

## 1-4. Transport measurements

For electrical transport measurements, samples were examined using a Physical Property Measurement System (Quantum Design), a $^4$He variable temperature insert equipped with a 15-Tesla superconducting magnet (Oxford Instruments plc.), or a 25 T Cryogen-free Superconducting Magnet (CSM) installed at High Field Laboratory for Superconducting Materials of Institute for Materials Research (IMR), Tohoku

University, Japan [4]. The longitudinal sheet resistance and Hall resistance were measured using standard lock-in technique with a typical excitation current of 10 nA and a repetition rate of 13 Hz. Electrical contacts were made using an indium solder. To exclude geometric effects, the measured data were symmetrized for $R_{xx}$ and anti-symmetrized for $R_{yx}$ against $B$.

## 2. Hall resistance of Fe$_x$(Bi$_{0.33}$Sb$_{0.67}$)$_{2-x}$Se$_3$/Bi$_2$Se$_3$ bilayers with various Fe content $x$.

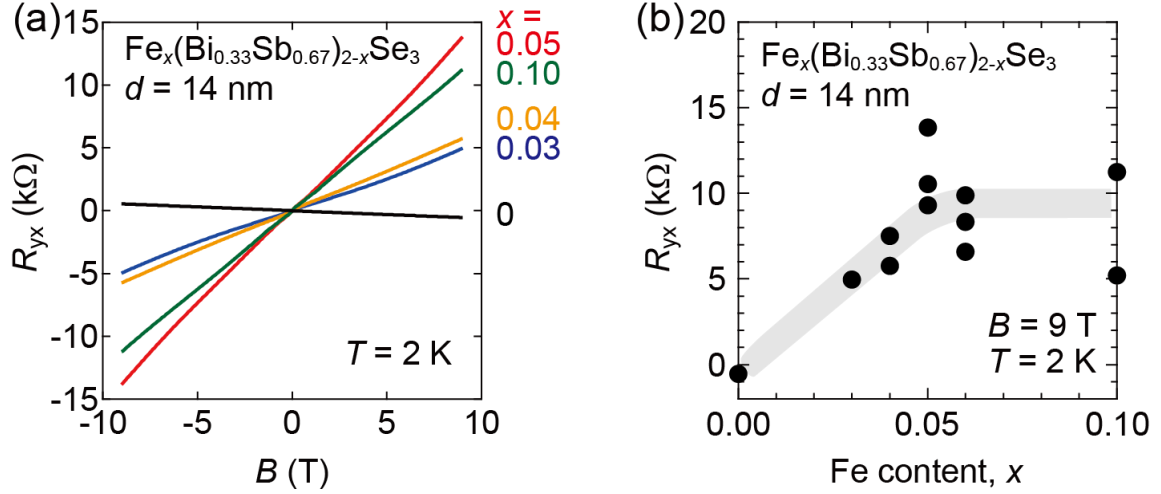

FIG. S1. Hall resistance of Fe$_x$(Bi$_{0.33}$Sb$_{0.67}$)$_{2-x}$Se$_3$/Bi$_2$Se$_3$ bilayers with various Fe contents $x$ and magnetic layer thickness $d = 14$ nm. (a) Hall resistance $R_{yx}$ as a function of the magnetic field $B$ at 2 K of Fe$_x$(Bi$_{0.33}$Sb$_{0.67}$)$_{2-x}$Se$_3$/Bi$_2$Se$_3$ bilayers with $x = 0$ (black), 0.03 (blue), 0.04 (yellow), 0.05 (red), and 0.10 (green). (b) $R_{yx}$ at 2 K and 9 T for Fe$_x$(Bi$_{0.33}$Sb$_{0.67}$)$_{2-x}$Se$_3$/Bi$_2$Se$_3$ bilayers as a function of Fe content $x$.

Figure S1(a) presents results of Hall effect measurements at $T = 2$ K for Fe$_x$(Bi$_{0.33}$Sb$_{0.67}$)$_{2-x}$Se$_3$/Bi$_2$Se$_3$ bilayers with $x = 0$, 0.03, 0.04, 0.05, and 0.10 and magnetic layer thickness of $d = 14$ nm. The Bi$_2$Se$_3$ buffer layer thickness is 3 nm for all studied samples. The top layer thickness of $d = 14$ nm is chosen to maximize $R_{yx}$ for Fe$_x$(Bi$_{0.33}$Sb$_{0.67}$)$_{2-x}$Se$_3$ with $x = 0.05$. Large positive slopes of $R_{yx}(B)$ are observed in Fe-doped samples, which constitute evidence that a large anomalous Hall term is generated by Fe doping. Figure S1(b) displays $R_{yx}$ at $T = 2$ K and in $B = 9$ T as a function of Fe content $x$. With increasing Fe content, $R_{yx}$ in $B = 9$ T tends to increase linearly up to $x = 0.05$ and to saturate for larger $x$. This saturation suggests the solubility limit of Fe.

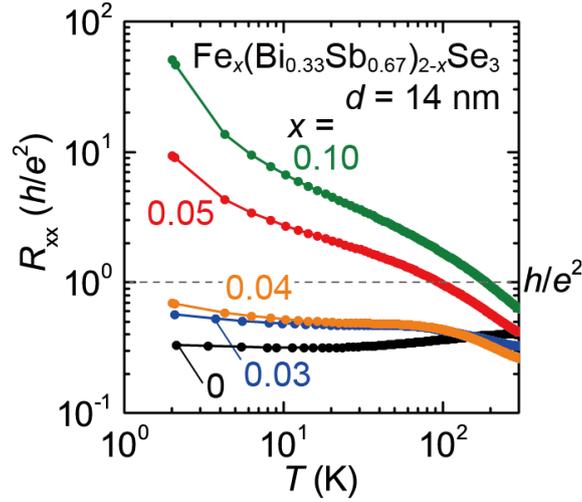

Figure S2 | Temperature dependence of the sheet resistance in $Fe_x(Bi_{0.33}Sb_{0.67})_{2-x}Se_3/Bi_2Se_3$ bilayer for various Fe content $x$.

Figure S2 shows corresponding temperature dependence of the sheet resistance $R_{xx}(T)$. With increasing Fe concentration, the metallic behavior for $x = 0$ turns into an insulating behavior in $x > 0.05$. Therefore, for 14-nm-thick $Fe_x(Bi_{0.33}Sb_{0.67})_{2-x}Se_3$/3-nm $Bi_2Se_3$ bilayer structure, the critical Fe concentration causing the Anderson insulator seems to be $x = 0.05$.

## 3. Electronic band structure of $Fe_{0.05}(Bi_{0.34}Sb_{0.66})_{1.95}Se_3$ from ARPES.

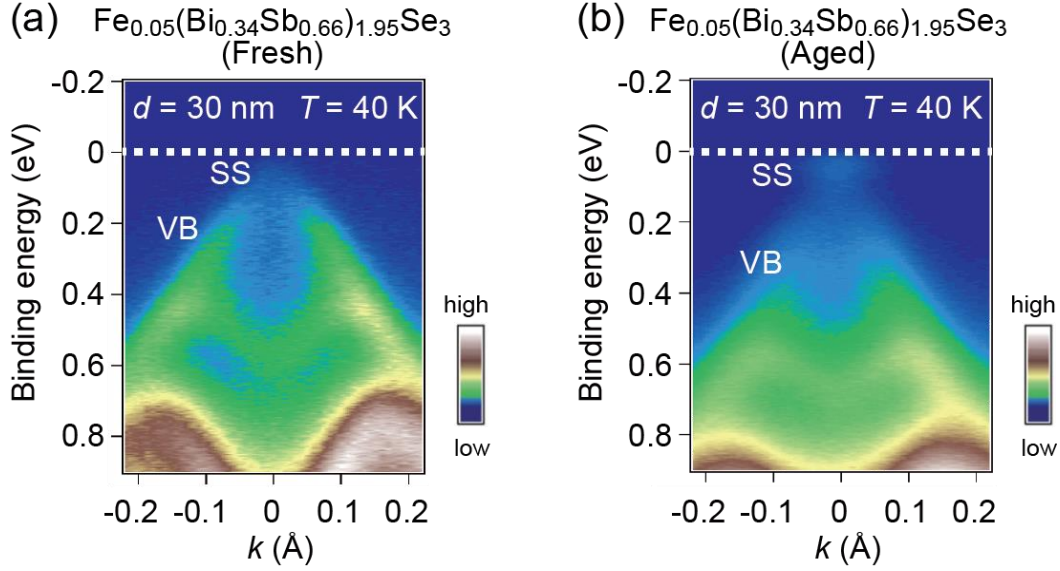

FIG. S3. Band structure of $Fe_{0.05}(Bi_{0.34}Sb_{0.66})_{1.95}Se_3$ from ARPES measurements. (a,b) ARPES results for the 30-nm-thick $Fe_{0.05}(Bi_{0.34}Sb_{0.66})_{1.95}Se_3$ layer on top of the $Bi_2Se_3$ buffer layer around the $\bar{\Gamma}$ point (a) before and (b) after surface aging. SS and VB respectively represent surface and bulk valence bands.

Figure S3(a) (corresponding to Fig. 1(g) in the main text) shows the electronic band structure along the $\bar{\Gamma} - \bar{K}$ cut of a $Fe_{0.05}(Bi_{0.34}Sb_{0.66})_{1.95}Se_3/Bi_2Se_3$ bilayer with magnetic layer thickness of $d = 30$ nm. The Fermi level is located slightly above the valence band. To study the surface band further, we have used the surface aging technique that increases band bending and shifts the Fermi level up, as shown in Fig. S3(b). In this manner, we visualize the linear dispersion and the Dirac point in the surface band. No gap at the Dirac point is detectable at this energy resolution, top layer thickness, and Fe concentration. In addition, compared to the data for a $(Bi,Sb)_2Se_3$ film [1], no discernible differences are detectable in the band structure and surface dispersion.

## 4. Temperature and magnetic-field dependence of transport properties for various $Fe_{0.05}(Bi_{0.34}Sb_{0.66})_{1.95}Se_3$ layer thicknesses.

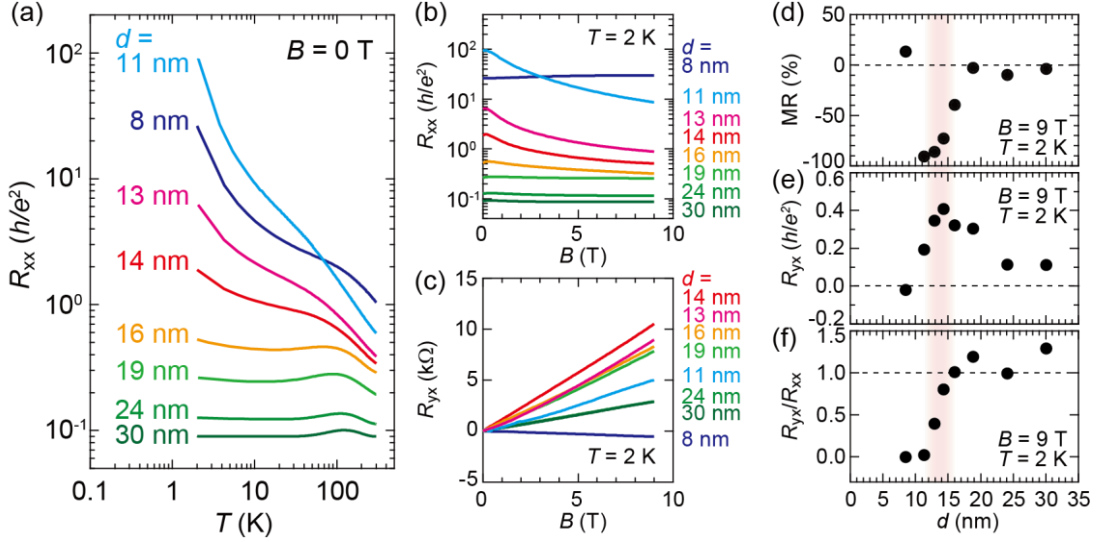

FIG. S4. Temperature and magnetic-field dependence of transport properties for $Fe_{0.05}(Bi_{0.34}Sb_{0.66})_{1.95}Se_3/Bi_2Se_3$ bilayers with various magnetic layer thickness $d$. (a) $R_{xx}(T)$ measured in $B = 0$, and (b) $R_{xx}(B)$ and (c) $R_{yx}(B)$ measured at $T = 2$ K for various magnetic layer thicknesses $d$ of 8 nm (blue) to 30 nm (green). Thickness dependences of (d) magneto-resistance ratio [defined as MR = $R_{xx}(B)/R_{xx}(0) - 1$], (e) $R_{yx}$ and (f) the Hall angle $R_{yx}/R_{xx}$ in $B = 9$ T and at $T = 2$ K.

Figure S4(a) shows the raw dataset of $R_{xx}(T)$, as presented in Fig. 1(h) in the main text for $Fe_{0.05}(Bi_{0.34}Sb_{0.66})_{1.95}Se_3/Bi_2Se_3$ bilayers in $B = 0$ T with the magnetic layer thickness $d$ varying from 8 nm (solid blue line) to 30 nm (solid green line). A metal-to-insulator transition occurs with decreasing $d$. Figures S4(b) and S4(c) present raw datasets $R_{xx}(B)$ and $R_{yx}(B)$ taken at 2 K, presented in Fig. 1(i) in the main text, for $Fe_{0.05}(Bi_{0.34}Sb_{0.66})_{1.95}Se_3/Bi_2Se_3$ bilayers with the magnetic layer thicknesses of 8 nm to 30 nm. Magnetoresistance ratio MR, $R_{yx}$ and the Hall angle $R_{yx}/R_{xx}$ in $B = 9$ T are presented respectively in Figs. S4(d) – S4(f). Magnitudes of both negative MR and positive $R_{yx}$ attain maximum values around $d = 14$ nm, as shown by the red shaded area. Note that a small positive MR and negative $R_{yx}$ for $d = 8$ nm originates from a remanent bulk conduction in bottom $n$-type $Bi_2Se_3$ where bulk carriers are not fully depleted by top $p$-type $Fe_{0.05}(Bi_{0.34}Sb_{0.66})_{1.95}Se_3/Bi_2Se_3$ layer [5].

## 5. Hybridization effects in non-magnetic and magnetic topological heterostructures.

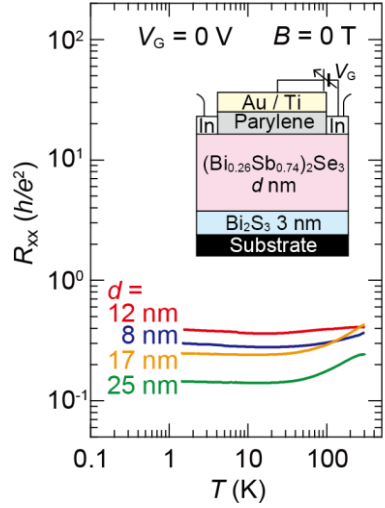

FIG. S5. Temperature dependence of sheet resistance $R_{xx}(T)$ of field-effect transistors based on $(Bi_{0.26}Sb_{0.74})_2Se_3/ Bi_2Se_3$ bilayer. Sheet resistance $R_{xx}(T)$ in a zero magnetic field and gate voltage for field effect transistors consisting of a 3 nm thick $Bi_2Se_3$ buffer layer and non-magnetic $(Bi_{0.26}Sb_{0.74})_2Se_3$ layers with thicknesses $d = 8$ (blue), 12 (red), 17 (orange), and 25 nm (green) [5].

### 5-1. $(Bi_{0.26}Sb_{0.74})_2Se_3/Bi_2Se_3$ bilayers

Because of hybridization between top and bottom surfaces, a gap opens at the Dirac point in TI films of the thicknesses below a critical value [6-11]. Figure S5 presents $R_{xx}(T)$ in $B = 0$ and at $V_G = 0$ for field-effect transistors of $(Bi_{0.26}Sb_{0.74})_2Se_3/Bi_2Se_3$ bilayers with the non-magnetic $(Bi_{0.26}Sb_{0.74})_2Se_3$ layer thickness varying from 8 (blue) to 25 nm (green). Data are extracted from our earlier report [5].

All devices show metallic behavior indicating that hybridization effects are not appreciable down to the 8 nm thickness in non-Fe-doped heterostructures even if the Fermi level is close to the neutrality point.

### 5-2. Earlier studies of $Bi_2Se_3$, $Bi_2Te_3$, $Sb_2Te_3$, and $(Bi,Sb)_2Te_3$ films

The critical thicknesses found using angle-resolved photoemission spectroscopy are approximately five quintuple layers (QL) [6] (one QL corresponds to approximately 1 nm), 1 QL [7], and 4 QL [9], respectively, for $Bi_2Se_3$, $Bi_2Te_3$, and $Sb_2Te_3$. For a $(Bi,Sb)_2Te_3$ thin film, the critical thickness is apparently less than 5 QL, because no gap is formed at this thickness [12].

### 5-3. 3$d$ transition metal-doped $Bi_2Se_3$ and $(Bi,Sb)_2Te_3$ films

Reportedly, the respective critical thicknesses of the Cu-doped $Bi_2Se_3$ and the Cr-doped $(Bi,Sb)_2Te_3$-based heterostructures are approximately 10 nm [13] and 13 nm [14]. These values are larger than the critical thicknesses of $Bi_2Se_3$ and $(Bi,Sb)_2Te_3$, respectively. The penetration depths of surface states $\xi$ are related directly to surface hybridization according to $\xi = \hbar v_F/|M|$, where $v_F$ is the Fermi velocity and $|M|$ is the charge excitation gap [15]. The $\xi$ magnitude can increase by doping with light elements, which decreases the spin-orbit interaction and consequently reduces $|M|$ [16,17]. A larger value of the critical thickness might derive from reduction of the spin-orbit interaction by 3$d$ transition metal doping. Therefore, it is plausible that the Fe-doped $(Bi,Sb)_2Se_3$ layer critical thickness is enhanced over the values reported for non-magnetic TIs. We present information for the critical thicknesses of various compounds in Table S1.

| Compounds | Critical thickness | Method | Reference |
|---|---|---|---|
| $Bi_2Se_3$ | 5 QL | Spectroscopy | ref. 6 |
| $Bi_2Te_3$ | 1 QL | Spectroscopy | ref. 7 |
| $Sb_2Te_3$ | 3 QL | Spectroscopy | ref. 9 |
| $(Bi,Sb)_2Te_3$ | < 5 QL | Spectroscopy | ref. 12 |
| Cu-doped $Bi_2Se_3$ | > 10 QL | Transport | ref. 13 |
| Cr-doped $(Bi,Sb)_2Te_3$ based heterostructure | > 13 QL | Transport | ref. 14 |

Table S1. Collection of critical thicknesses at which top and bottom surface states hybridize for various compounds.

## 6. Magneto-transport properties in perpendicular and in-plane magnetic fields.

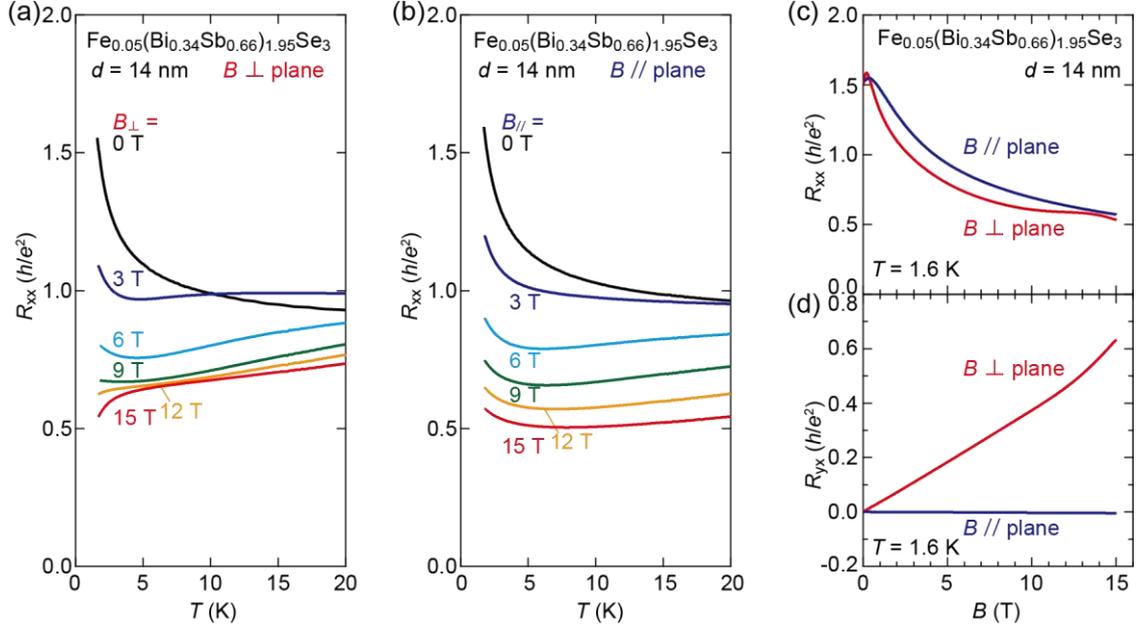

FIG. S6. Role of the magnetic field orientation. (a,b) $R_{xx}(T)$ dependencies for a $Fe_{0.05}(Bi_{0.34}Sb_{0.66})_{1.95}Se_3/Bi_2Se_3$ bilayer with the magnetic layer thickness $d = 14$ nm in various magnetic fields $B$. The magnetic field is applied (a) perpendicularly to the sample plane and (b) in-plane. (c,d) $R_{xx}(B)$ and $R_{yx}(B)$ dependencies in perpendicular (red curves) and in-plane (blue curves) magnetic fields at $T = 1.6$ K for the same sample.

To explore characteristic of the field-induced topological phase transitions further, we studied transport properties for two orientations of the magnetic field with respect to the film plane. Figure S6 shows $R_{xx}(T)$ in perpendicular (Fig. S6(a)) and in-plane (Fig. S6(b)) magnetic fields for a $Fe_{0.05}(Bi_{0.34}Sb_{0.66})_{1.95}Se_3/Bi_2Se_3$ bilayer with magnetic layer thickness $d = 14$ nm. A similar magnitude of negative magnetoresistance is observed for the two field directions in the high-temperature region, substantiating the role of the magnetic field in reducing scattering by Fe spins. At low temperatures, however, the insulator-to-metal transition is observed only by application of an out-of-plane magnetic field, whereas the insulating behaviour [$dR_{xx}(T)/dT < 0$] remains under the in-plane magnetic field of 15 T. Therefore, metallization is observed only when the magnetic field is applied perpendicular to the film plane, which substantiates our claim that the quantum phase transition occurs from the 2D-insulator to the quantum anomalous Hall state.

## 7. $R_{xx}(B)$ and $R_{yx}(B)$ measured for the FET device for various gate voltages $V_G$.

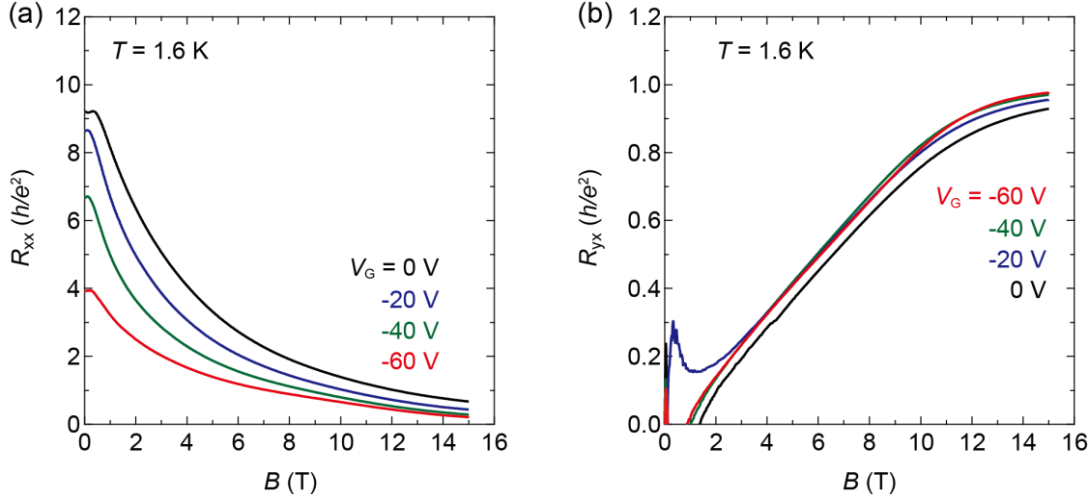

FIG. S7. $R_{xx}(B)$ and $R_{yx}(B)$ of the FET device of the trilayer structure under various $V_G$. (a,b) Magnetic-field dependences of (a) $R_{xx}$ and (b) $R_{yx}$ measured for the FET device based on trilayer at $T = 1.6$ K under various $V_G$.

Figure S7 shows magnetic-field dependence of $R_{xx}$ and $R_{yx}$ for the FET device based on the trilayer (the inset of Fig. 3(b) in the main text) at $T = 1.6$ K under $V_G = 0$ V (black), −20 V (blue), −40 V (green), and −60 V (red). At zero magnetic field, applying a negative $V_G$ decreases $R_{xx}$. By sweeping $B$ in the negative $V_G$, longitudinal resistance $R_{xx}$ decreases towards zero and Hall resistance $R_{yx}$ almost reaches quantized resistance $h/e^2$.

## 8. $R_{xx}(T)$ and $R_{yx}(T)$ measured for the FET device under various magnetic fields.

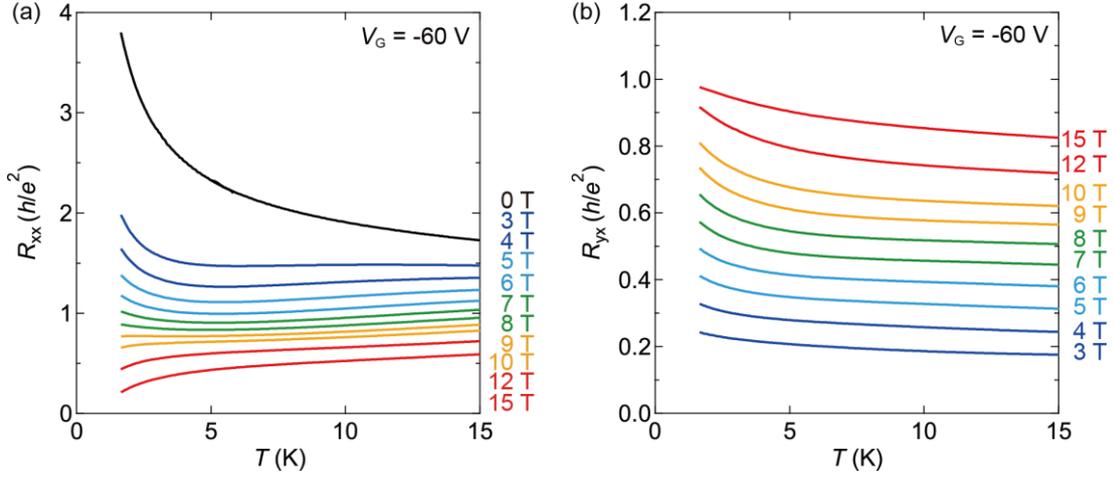

FIG. S8. (a,b) Temperature dependences of (a) $R_{xx}$ and (b) $R_{yx}$ of the FET device of the trilayer structure with $V_G$ = -60 V under various $B$.

Figure S8 shows temperature dependences of $R_{xx}$ and $R_{yx}$ for the FET device based on the trilayer (the inset of Fig. 3(b) in the main text) at $V_G$ = -60 V under various perpendicular magnetic field $B$ from 0 T (black) to 15 T (red). At zero magnetic field, $R_{xx}$(T) shows insulating behavior. By increasing $B$, longitudinal resistance $R_{xx}$ decreases and becomes metallic and Hall resistance $R_{yx}$ increases and almost reaches quantized resistance $h/e^2$ under $B$ = 15 T at $T$ = 1.6 K. At low field, $R_{yx}(T)$ cannot properly measured owing to high longitudinal resistance.

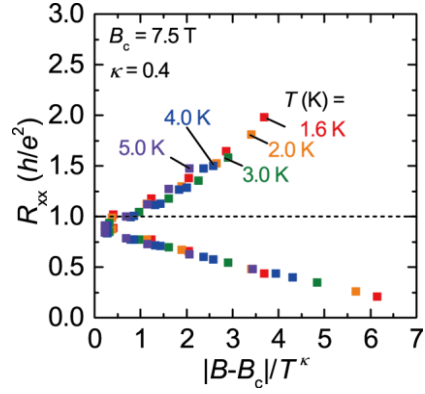

Figure S9 | $R_{xx}$ plotted against the scaled magnetic field $|B-B_c/T^\kappa|$ at different temperature $T$. The data are extracted from $T$ dependence of $R_{xx}$ measured at different $B$ presented in Fig. S8.

We used the scaling relation that $R_{xx}$ is a function of $|B-B_c/T^\kappa|$ to scale $R_{xx}(T)$ measured at different magnetic field presented in Fig. S8. Figure S9 shows $R_{xx}$ at typical temperatures ($T$ = 1.6, 2.0, 3.0, 4.0, and 5.0 K) are plotted in $|B-B_c/T^\kappa|$ by substituting 7.5 T and 0.4 into the critical magnetic field $B_c$ and the critical exponent $\kappa$. The curves at different $T$ converge to a single curve around the critical point. The critical magnetic field $B_c \sim 7.5$ T is consistent with renormalization group flow of the conductivity tensor discussed in main Fig. 4.

# 9. Estimation of energies accounting for band crossing in Fe-doped (Bi,Sb)$_2$Se$_3$/Bi$_2$Se$_3$ bilayer.

It is anticipated that the QAH phase can occur if $\Delta_{Zeeman}^2 > V^2 + \Delta_{hy}^2$, where $V$ is the energy difference between the positions of the Dirac point at the top and bottom surfaces [18,19]. In addition, the scattering broadening of surface states $\gamma \cong \hbar/\tau$ must be considered. Here, $\hbar$ and $\tau$ are the reduced Planck's constant and the electron lifetime, usually shorter than the momentum relaxation time. We estimate that it is below $\gamma \cong 40$ meV. Zeeman energy of $\Delta_{Zeeman} = g_s \mu_B B$ with $g_s$ and $\mu_B$ being effective g-factor and Bohr magneton, is estimated as approximately 16–43 meV in $B = 15$ T when using the reported value of the g-factor $g_s = 18$–50 for the surface state of Bi$_2$Se$_3$ [20,21]. Actually, because of *sp-d* exchange interactions between band carriers and spins localized on the *d* shells of magnetic ions, the Zeeman splitting can be enhanced in semiconductors containing transition-metal impurities [22-25]. It is anticipated that similar enhancement of the g-factor in Fe-doped Bi$_2$Se$_3$ engenders $\Delta_{Zeeman}$ larger than $V$ and $\gamma$. As explained below, the value of $V$ is estimated to be 80 meV. The typical value of $\Delta_{hy}$ in Bi$_2$Se$_3$ thin films is 0.1-0.2 eV [6]. Therefore, the g-factor that requires to fulfil the relation $\Delta_{Zeeman}^2 > V^2 + \Delta_{hy}^2$, is roughly 250 under the application of the magnetic field of 15 T.

## 9-1. Potential variation $V$ at top and bottom surface states

Energy difference between the Dirac points at the top and bottom surfaces corresponds to built-in potential of the *p-n* junction in the relevant Fe$_x$(Bi,Sb)$_2$Se$_3$/Bi$_2$Se$_3$ bilayer structure. From Poisson's equation for the semiconductor *p-n* junction, built-in potential ($V_{bi}$) is given as $V_{bi} = e/\varepsilon_r\varepsilon_0 (nd_n^2 + pd_p^2)$, where $e$, $\varepsilon_r$, and $\varepsilon_0$ respectively represent the elementary charge, dielectric constant, permittivity of vacuum, and where $n$, $p$, $d_n$, and $d_p$ respectively represent electron and hole carrier densities and the depletion layer thickness in the $n$ and $p$-type region. $V_{bi}$ in the bilayer system can be calculated as approximately 80 meV using parameters extracted from earlier reports $p = 5 \times 10^{18}$ cm$^{-3}$ and $d_p = 12$ nm for (Bi,Sb)$_2$Se$_3$/Bi$_2$Se$_3$ bilayer and $n = 2 \times 10^{19}$ cm$^{-3}$ and $d_n = 3$ nm for Bi$_2$Se$_3$ [5]. In addition, the dielectric constant of Bi$_2$Se$_3$ ($\varepsilon_r = 100$) is used for calculations [26].

## 9-2. Scattering broadening

Scattering broadening $\gamma$ of electronic states is of the order of $\hbar/\tau = \hbar e/\mu m_c$, where $\hbar$ and $\tau$ respectively represent the reduced Planck's constant and electron scattering time, and where $e$, $\mu$, and $m_c$ represent the elemental charge, electron mobility, and effective mass. Using the measured value of $\mu = 300$ cm$^2$V$^{-1}$s$^{-1}$ for non-Fe-doped (Bi,Sb)$_2$Se$_3$/Bi$_2$Se$_3$ bilayer [1] and the reported value of $m_c = 0.1\ m_0$ [21], $\gamma$ can be evaluated as approximately 40 meV.

## 9-3. *g*-factor enhancement in dilute magnetic semiconductors

A unique feature in dilute magnetic semiconductors is so-called giant Zeeman splitting, which results from *sp-d* exchange interactions [22,23]. Aligning local magnetic moments of magnetic impurities by an external magnetic field induces large Zeeman splitting in the conduction and valence bands via the *sp-d* exchange interaction. A typical value of the Zeeman splitting in Mn or Fe-doped II-VI semiconductors reaches a few tens of milli-electron volts in the presence of a magnetic field of a few Teslas and at low temperatures [24,25].

**9-4. Reported values for the gap energy and related values for magnetic (Bi,Sb)$_2$Te$_3$**

The size of the gap energy of surface states in a ferromagnetic Cr$_{0.08}$(Bi$_{0.1}$Sb$_{0.9}$)$_{1.92}$Te$_3$ was reported as approximately 30 meV using scanning tunneling microscopy [27]. Scattering broadening γ can be estimated as approximately 25 meV using reported values of $\mu$ ~ 500 cm$^2$V$^{-1}$s$^{-1}$ [9] and $m_c$ ~ 0.1 $m_0$ [28] for (Bi,Sb)$_2$Te$_3$.

**10. Gate voltage dependence of $R_{xx}(B)$ and $R_{yx}(B)$ in a field-effect transistor (FET) based on the trilayer structure.**

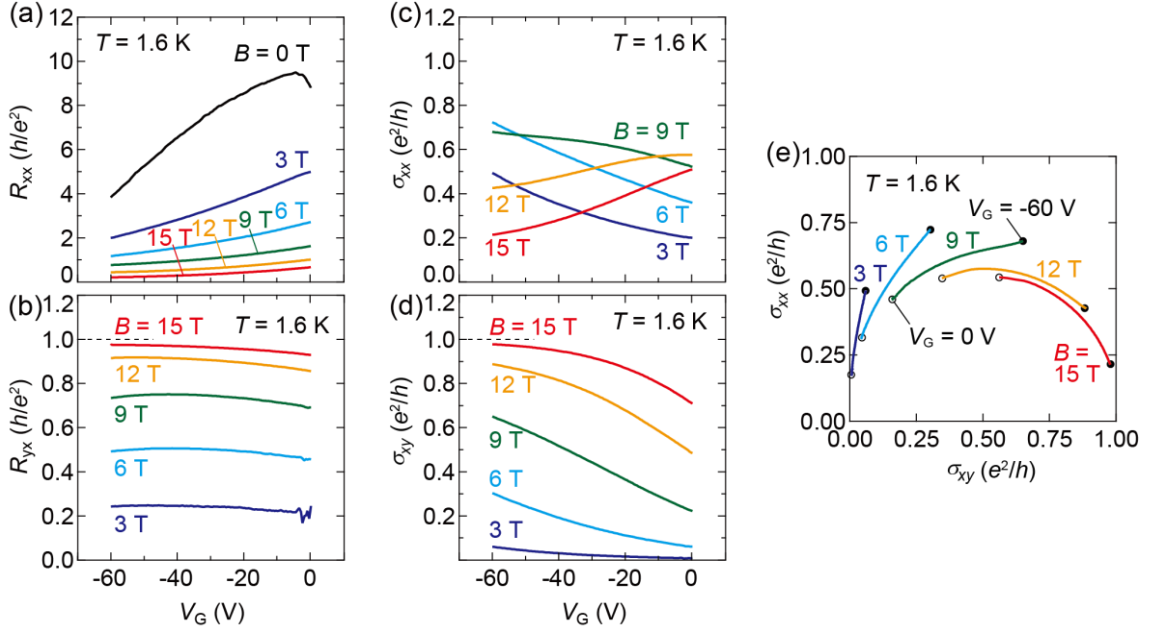

FIG. S10. Dependence of $R_{xx}$ and $R_{yx}$ on the gate voltage $V_G$ at various magnetic fields. (a–d) $V_G$ dependence of (a) longitudinal resistance $R_{xx}$; (b) Hall resistance $R_{yx}$; (c) longitudinal conductivity $\sigma_{xx}$ and (d) Hall conductivity $\sigma_{xy}$ under various magnetic fields $B = 0$ (black), 3 (blue), 6 (cyan), 9 (green), 12 (orange), and 15 T (solid red line) at $T = 1.6$ K in the field-effect transistor device presented in Figs. 3(b) and 4 in the main text. (e) $\sigma_{xx}(V_G)$, $\sigma_{xy}(V_G)$ plot at $T = 1.6$ K for various magnetic fields. Empty and filled circles represent data at $V_G = 0$ V and -60 V.

Figure S10 presents dependence of $R_{xx}$ and $R_{yx}$ on the gate voltage $V_G$ in the field-effect transistor device consisting of a 1-nm-thick $Bi_2Se_3$/18.5-nm-thick $Fe_{0.05}(Bi_{0.33}Sb_{0.67})_{1.95}Se_3$/3-nm-thick $Bi_2Se_3$ tri-layer structure (inset to Fig. 3(b)). Under $B = 15$ T, $\sigma_{xx}$ and $\sigma_{xy}$ converge to $(\sigma_{xx}, \sigma_{xy}) = (0, e^2/h)$ of the quantum anomalous phase by sweeping $V_G$ to the negative value, as shown in Fig. S10(e). The result indicates that the Fermi level approaches the magnetic-field-induced gap at the Dirac point by negative $V_G$. Moreover, at $B$ = 9 and 12 T, the data point of ($\sigma_{xx}$, $\sigma_{xy}$) moves from the trivial region $\sigma_{xy} < 0.5\ e^2/h$ to QAH region by application of $V_G$. Detailed measurement of temperature dependence is necessary to ascertain the ground state; we leave this assessment as a subject for future study.

# 11. Structural analysis

Figure S11 presents x-ray diffraction patterns and atomic force microscopic images of 15-nm-thick Fe-doped $(Bi,Sb)_2Se_3$ layers with Fe content up to $x = 0.2$. For $x = 0$ (bottom panel), a clear diffraction peak (solid triangle) is observed at the $2\theta$ angle of 18.6°, which is assigned to $(Bi,Sb)_2Se_3$ (0006). At the same time, a step-and-terrace structure is observed using AFM, as reported already [1]. For $x = 0.05$, an identical diffraction pattern and step-and-terrace structure are found. When $x$ is increased to 0.10, rectangular precipitates start to appear on the surface. For $x = 0.20$, the number of such precipitates increases. An additional x-ray diffraction peak appears at around $2\theta$ angle of 16.1°, which we assigned to $Fe_3Se_4$ (001) [29]. Based on the results of analyses presented above, we evaluate the solubility limit of Fe as approximately $x = 0.1$.

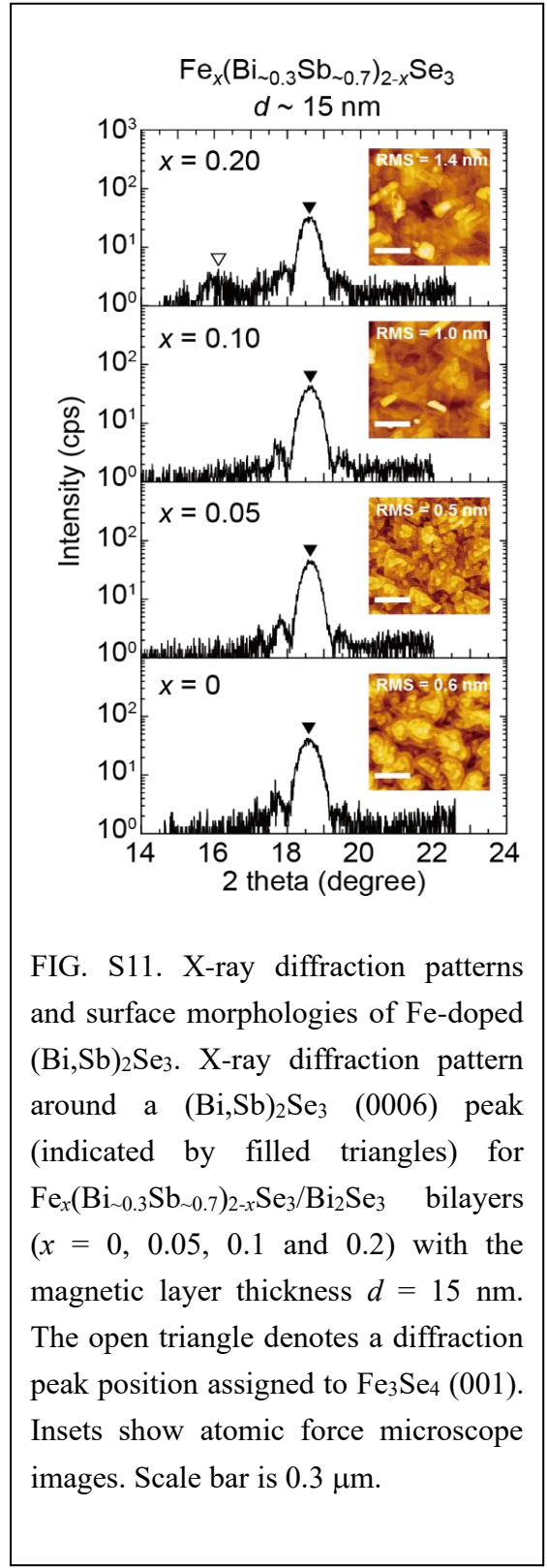

FIG. S11. X-ray diffraction patterns and surface morphologies of Fe-doped $(Bi,Sb)_2Se_3$. X-ray diffraction pattern around a $(Bi,Sb)_2Se_3$ (0006) peak (indicated by filled triangles) for $Fe_x(Bi_{\sim 0.3}Sb_{\sim 0.7})_{2-x}Se_3/Bi_2Se_3$ bilayers ($x = 0, 0.05, 0.1$ and $0.2$) with the magnetic layer thickness $d = 15$ nm. The open triangle denotes a diffraction peak position assigned to $Fe_3Se_4$ (001). Insets show atomic force microscope images. Scale bar is 0.3 μm.

embedded in Bi$_2$Se$_3$ topological insulator thin films grown by molecular beam epitaxy, ACS Nano **10**, 1132 (2016).